\documentclass[12 pt]{article}
\usepackage{graphicx} 
\usepackage{natbib}
\usepackage{amsthm}
\def\bSig\mathbf{\Sigma}

\newtheorem{proposition}{Proposition}

\newtheorem{remark}{Remark}

\title{A novel smoothing-based goodness-of-fit test of covariance for multivariate sparse functional data}

\usepackage{amsmath,amsfonts,amssymb,booktabs,color,epsfig,graphicx,hyperref,url}

\usepackage{enumitem}
\newlist{subquestion}{enumerate}{1}
\setlist[subquestion,1]{label=(\alph*)}
\usepackage{graphicx}
\usepackage{float}
\usepackage{multirow}
\usepackage{multicol}
\usepackage{amssymb}
\usepackage{calrsfs}
\usepackage[mathscr]{eucal}
\usepackage{amsmath}
\usepackage{xcolor}

\usepackage{hyperref}  
\hypersetup{
    colorlinks=true,
    linkcolor=blue,
    filecolor=magenta,      
    urlcolor=brown,
    citecolor = red,
    pdftitle={Overleaf Example},
    pdfpagemode=FullScreen,
    }
\usepackage{url}            
\usepackage{booktabs}       
\usepackage{amsfonts}       
\usepackage{nicefrac}       
\usepackage{bm}
\usepackage[width = 7in, height = 9in]{geometry}
\usepackage{tikz}
\usetikzlibrary{arrows.meta}
\usepackage{verbatimbox}
\usepackage{listings}

\usepackage{blindtext}
\usepackage{tabularx}
\usepackage{algorithm2e}
\usepackage{algpseudocode}

\author
{Dhrubajyoti Ghosh
\\
Department of Biostatistics and Bioinformatics, 
\\
Zhuolin Song
\\
Department of Statistics, North Carolina State University, 
\\
Luo Xiao
\\
Department of Statistics, North Carolina State University, 
\\
Sheng Luo
\\
Department of Biostatistics and Bioinformatics.
}
\providecommand{\keywords}[1]{\textbf{\textit{Keywords---}} #1}

\begin{document}
\maketitle

\begin{abstract}
Accurately specifying covariance structures is critical for valid inference in longitudinal and functional data analysis, particularly when data are sparsely observed. In this study, we develop a global goodness-of-fit test to assess parametric covariance structures in multivariate sparse functional data. Our contribution is twofold. First, we extend the univariate goodness-of-fit test proposed by \cite{chen2019smoothing} 
to better accommodate sparse data by improving error variance estimation and applying positive semi-definite smoothing to covariance estimation. These corrections ensure appropriate Type I error control under sparse designs. Second, we introduce a multivariate extension of the improved test that jointly evaluates covariance structures across multiple outcomes, employing novel test statistics based on the maximum and $\ell_2$ norms to account for inter-outcome dependencies and enhance statistical power. Through extensive simulation studies, we demonstrate that the proposed methods maintain proper Type I error rates and achieve greater power than univariate tests with multiple testing adjustments. Applications to longitudinal neuroimaging and clinical data from the Alzheimer's Disease Neuroimaging Initiative (ADNI) and the Parkinson's Progression Marker Initiative (PPMI) illustrate the practical utility of the proposed methods for evaluating covariance structures in sparse multivariate longitudinal data.
\end{abstract}



\section{Introduction}\label{sec:intro}

Neurodegenerative disorders, such as Alzheimer's disease (AD) and Parkinson's disease (PD), are progressive conditions that impair cognitive and motor functions through the gradual loss of neurons in critical brain regions. To understand the complex mechanisms underlying these diseases and to track their progression, researchers often rely on sparse multivariate longitudinal data, where multiple outcomes such as clinical scores, imaging measures, and biomarkers are repeatedly observed over time. However, in many real-world studies, these data are collected irregularly and infrequently, with subjects typically having only a few observations at varying time points. Such data structures pose significant challenges for statistical modeling, particularly in accurately capturing the covariance structures among outcomes over time.

Traditional approaches for analyzing multivariate longitudinal data include linear mixed-effects models and generalized estimating equations (GEE) \citep{verbeke1997linear,fitzmaurice2004applied}. These models are effective at capturing both population-level trajectories and subject-specific variability, and they allow for correlations among outcomes. However, they often rely on pre-specified parametric covariance structures such as autoregressive or compound symmetry forms. Such assumptions may be overly simplistic or incorrect in practice, especially in complex real-world data with sparse observations. Extensions such as joint models have been proposed to study longitudinal outcomes in relation to terminal events \citep{wang2017dynamic}, but they still depend on assumed covariance structures that may not hold in practice.

An alternative framework is functional data analysis (FDA), which treats longitudinal measurements as continuous functions over time \citep{ramsay2005functional, guo2002functional}. In the multivariate setting, multivariate functional data analysis (MFDA) enables joint modeling of multiple correlated functions, and multivariate functional principal component analysis (MFPCA) can reduce dimensionality while preserving dependencies among outcomes \citep{happ2018multivariate}. However, FDA methods face practical limitations, especially when applied to sparse functional data, where only a few time points are observed per subject. Basis function selection, computational burden, and instability in estimation are additional challenges. To address these, \cite{li2020fast} developed multivariate Fast Covariance Estimation (mFACEs), a method designed specifically for efficient and accurate estimation of covariance structures in sparse multivariate functional data.

Regardless of the modeling framework, accurate specification and validation of covariance structures is essential for reliable inference and prediction. While many tests for covariance structures have been developed in high-dimensional settings \citep{ledoit2002some, bai2009corrections, chen2010tests}, they typically assume regular observation designs and are not suited to the smooth covariance structures inherent in functional data. \cite{zhong2017tests} proposed a general goodness-of-fit (GOF) test for parametric covariance structures under fixed designs, but this approach is not applicable when observation times vary across subjects. More recently, \cite{chen2019smoothing} introduced a smoothing-based GOF test for univariate functional data under both fixed and random designs, incorporating smoothness in covariance estimation. However, this method fails to control Type I error when data are sparse, due to biased error variance estimation. Moreover, its extension to multivariate data relies on multiple testing corrections such as Bonferroni adjustment, which reduce power and ignore correlations among outcomes \citep{li2023latent}.

To address these limitations, we propose a global goodness-of-fit test for covariance structures in sparse multivariate functional data, with two major contributions. First, we propose an improved univariate goodness-of-fit (GOF) test of covariance that extends the original univariate GOF test by \cite{chen2019smoothing} to better accommodate sparse functional data. This is achieved by incorporating an enhanced error variance estimation via Fast Covariance Estimation (FACE) \citep{xiao2018fast} and by applying positive semi-definite smoothing to ensure valid covariance estimation. This improvement restores Type I error control in sparse settings where the original test fails. Second, we develop a multivariate extension of the improved GOF test that evaluates the covariance structures across multiple outcomes jointly, using novel test statistics based on max and $\ell_2$ norms. These statistics enhance power by aggregating deviations across outcomes while accounting for dependencies among outcomes, thereby avoiding the inefficiencies of separate univariate tests with multiple testing corrections. We conduct extensive simulation studies to evaluate the proposed methods under realistic sparse designs. The results demonstrate robust Type I error control and improved power compared to existing approaches. We further illustrate the practical value of our methods by applying them to data from two major neurodegenerative disease studies: the Alzheimer's Disease Neuroimaging Initiative (ADNI) and the Parkinson's Progression Marker Initiative (PPMI). These analyses highlight that accurately testing covariance structures not only improves understanding of disease progression but also guides appropriate model selection in longitudinal data analysis.

The remainder of this paper is organized as follows. Section \ref{sec:method} presents the statistical framework, reviews the original univariate GOF test \citep{chen2019smoothing}, describes our improvements for sparse data, and details the proposed multivariate extension. Section \ref{sec:simul} reports simulation studies evaluating Type I error control and power. Section \ref{sec:realData} applies our methods to real-world datasets from ADNI and PPMI. Section \ref{sec:concl} concludes with a discussion of implications and future research directions.

\section{Methodology} \label{sec:method}

Our methodology builds on the original univariate GOF test \citep{chen2019smoothing}, a foundational tool for assessing parametric covariance structures in functional data. In Section 2.1, we summarize this original univariate test and its core components. In Section 2.2, we introduce two key improvements that address its limitations for sparse functional data: refined error variance estimation  and a truncation procedure to ensure positive semi-definiteness of estimated covariance functions. These modifications restore control of the Type I error rate and validity of the original test under sparsity. Section 2.3 presents the multivariate extension to jointly evaluate covariance structures across multiple longitudinal outcomes. This multivariate GOF test (MGFC) is designed for multivariate sparse functional data and offers improved power and reliable Type I error control in practical applications.

\subsection{Review of Univariate Goodness-of-Fit Test of Covariance}
\label{subsec:univariate}
Let $(t_{ij}, Y_{ij}) \in \mathcal{T} \times \mathbb{R}: i=1,\ldots,N, j=1, \ldots, J_i$ denote functional or longitudinal observations, where $i$ indexes subjects, $j$ indexes visit numbers (with $J_i$ varying across subjects), and $Y_{ij}$ is the measurement for subject $i$ at time $t_{ij}$. The data are modeled as
\begin{equation}
    Y_{ij} = \mu(t_{ij}) + X_i(t_{ij}) + \epsilon_{ij}, \quad 1 \leq i \leq N, \; 1 \leq j \leq J_i,
    \label{basicModel}
\end{equation}
where $\mu(\cdot)$ is a smooth mean function, $\{X_i\}$ are independent zero-mean Gaussian random functions, and $\epsilon_{ij}$ are independent identically distributed Gaussian errors with mean zero and variance $\sigma^2$, independent of $X_i$.

Let $C(t, t') = \text{Cov}(X_i(t), X_i(t'))$ denote the covariance function of $X_i(t)$, assumed to be a smooth and positive semi-definite function on $\mathcal{T}^2$. In longitudinal data analysis, a parametric form of $C(t, t')$ is often specified, whereas functional data analysis typically assumes only smoothness and flexibility. It is of interest to test whether the true covariance equals a specified parametric form $C_0(t, t')$, via the hypothesis:
\begin{equation}
  H_0: C(t,t') = C_0(t,t'), \; \forall (t, t')\in\mathcal{T}^2 \quad \text{vs} \quad H_A: C(t,t') \neq C_0(t,t') \text{ for some } (t, t')\in\mathcal{T}^2 .
\label{eq:generalHypo}
\end{equation}
This hypothesis test formulated by \cite{chen2019smoothing} is versatile and applicable to various smooth $C_0$.
 And a common choice for $C_0(t, t')$ arises from the random linear mixed effects model: $X_i(t) = b_{0i} + b_{1i} t$,
where $\textbf{b}_i = (b_{0i}, b_{1i})^T$ are random effects with zero mean and covariance $\text{Var}(b_{0i}) = \sigma_0^2$, $\text{Var}(b_{1i}) = \sigma_1^2$, $\text{Cov}(b_{0i}, b_{1i}) = \sigma_{01}$. The implied covariance function is
\begin{equation}
C_0(t,t') = \sigma_0^2 + \sigma_{01}(t + t') + \sigma_1^2 tt',
\label{eq:quadform}
\end{equation}
with unknown parameters $(\sigma_0^2, \sigma_{01}, \sigma_1^2)$ to be estimated.

The mean function $\mu(t)$ is estimated using penalized splines \citep{ruppert2003semiparametric, wood2003thin}, yielding demeaned observations $\widetilde{Y}_{ij} = Y_{ij} - \widehat{\mu}(t_{ij})$. Under $H_0$ with $C_0$ as in \eqref{eq:quadform}, the variance parameters $(\sigma_0^2, \sigma_{01}, \sigma_1^2)$ are estimated by maximizing the likelihood, and the estimated null covariance is $\widehat{C}_0(t, t') = \widehat{\sigma}_0^2 + \widehat{\sigma}_{01}(t + t') + \widehat{\sigma}_1^2 tt'$. Under $H_A$, the covariance $C_A(t, t')$ is approximated using tensor-product regression splines: $C_A(t, t') = \sum_{h=1}^H \sum_{\ell=1}^H \theta_{h\ell} B_h(t) B_\ell(t')$, where $\{B_h(t)\}$ are cubic B-spline basis functions over $\mathcal{T}$. The coefficients $\theta_{h\ell}$ are estimated by minimizing the least squares criterion:
\begin{equation}
\sum_{i=1}^N \sum_{1 \leq j \neq j' \leq J_i} \left\{ \widetilde{Y}_{ij}\widetilde{Y}_{ij'} - \sum_{h,\ell = 1}^H \theta_{h\ell}B_h\left(t_{ij}\right)B_\ell\left(t_{ij'}\right) \right\}^2.
\label{eq:Bspline}
\end{equation}
Let $\widehat{\theta}_{h\ell}$ denote the minimizer; then the estimated alternative covariance function is $\widehat{C}_A(t,t') = \sum_{h,\ell=1}^H \widehat{\theta}_{h\ell} B_h(t) B_\ell(t')$.

The test statistic for evaluating the goodness-of-fit is $T_N = \lVert \widehat{C}_A - \mathcal{K}\widehat{C}_0 \rVert_{HS}$, where $\lVert \cdot \rVert_{HS}$ denotes the Hilbert-Schmidt norm, defined as $\lVert f \rVert_{HS} = \left[ \int\!\int f^2(t,t') \,dt\,dt' \right]^{1/2}$. The subscript $N$ indicates that the test statistics is computed based on data of sample size $N$. The operator $\mathcal{K}$ applies smoothing to $\widehat{C}_0$ by fitting a tensor-product spline to its values:
Specifically, $\widehat{C}_0(t_{ij}, t_{ij'})$ is substituted for $\widetilde{Y}_{ij} \widetilde{Y}_{ij'}$ in \eqref{eq:Bspline} to obtain coefficients $\widehat{\theta}_{0,h\ell}$, yielding $\mathcal{K}\widehat{C}_0(t,t') = \sum_{h,\ell=1}^H\widehat{\theta}_{0,h\ell}B_h(t)B_\ell(t')$. Smoothing $\widehat{C}_0$ ensures a fair comparison with $\widehat{C}_A$ and reduces smoothing bias \citep{hardle1993comparing}. The error variance $\sigma^2$ under $H_A$ is estimated by comparing diagonal elements of the sample covariance matrix to those of $\widehat{C}_A$, using methods from \citet{yao2005functional} and \citet{goldsmith2013corrected}.

To approximate the null distribution of $T_N$, \citet{chen2019smoothing} employ wild bootstrap, a resampling method common in nonparametric tests \citep{hardle1993comparing}, which uses random generated errors to generate replicates under $H_0$. Importantly, the variance of the random errors is estimated under the alternative.

\subsection{Improved Univariate Goodness-of-Fit Test of Covariance}
\label{subsec:face}

While the original univariate GOF test \citep{chen2019smoothing} performs well for densely sampled functional data, it exhibits inflated Type I error rates when applied to sparse functional data where each subject has only a few observations. This issue stems from the underestimation of the error variance $\sigma^2$ under the alternative, as tensor-product splines tend to overfit the covariance function in settings with limited data. Notably, the estimated $\sigma^2$ under the alternative model is used to generate random noise under the null hypothesis during wild bootstrap. As a result, underestimated $\sigma^2$ produces a null distribution for the test statistic with artificially reduced variability, yielding smaller p-values and inflated Type I errors.

This limitation makes the original univariate test unreliable for longitudinal data with limited observations per subject. To address the issue, we incorporate two improvements  into the univariate test. First, we estimate the error variance $\sigma^2$ using the fast covariance estimation for sparse functional data (FACEs) proposed by \cite{xiao2018fast}. FACEs estimates both the smooth covariance function and the error variance simultaneously using bivariate penalized splines, and has demonstrated superior accuracy in estimating $\sigma^2$ compared to competing methods such as PACE \citep{yao2005functional}, \textit{pffr} \citep{scheipl2015functional}, and FAMM \citep{cederbaum2016functional}. While FACEs requires smoothing parameter selection and is computationally more intensive than tensor-product regression splines, we use FACEs solely for estimating $\sigma^2$. In the wild bootstrap procedure, tensor-product splines remain in use for estimating the alternative covariance due to their computational efficiency. An alternative method for estimating $\sigma^2$ is SNPT \citep{Lin2022}, which performs similarly to FACEs \citep{Song2024}.

Second, tensor-product regression splines often yield covariance estimates with negative eigenvalues, requiring truncation to ensure positive semi-definiteness. However, naive truncation introduces bias. To mitigate this, we perform optimal truncation of the tensor-product spline estimator, ensuring positive semi-definiteness while minimizing the Hilbert--Schmidt distance to the initial estimate. Specifically, the covariance under the alternative is represented as
$C_A(t,t') = \mathbf{b}(t)^{\mathrm{T}} \mathbf{\Theta} \mathbf{b}(t')$, where $\mathbf{b}(t) = (B_1(t), \ldots, B_H(t))^{\mathrm{T}}$ is the vector of B-spline basis functions, and $\mathbf{\Theta} = (\theta_{h\ell}) \in \mathbb{R}^{H \times H}$ is a symmetric coefficient matrix. We establish two propositions to support this procedure.

\begin{proposition}
\label{prop:psd_iff}
The covariance function 
$C_A(t,t') = \mathbf{b}(t)^{\mathrm{T}} \boldsymbol{\Theta} \, \mathbf{b}(t')$ 
with symmetric $\boldsymbol{\Theta}$ is positive semi-definite if and only if $\boldsymbol{\Theta}$ is positive semi-definite.
\end{proposition}

Although truncating negative eigenvalues of the tensor-product spline estimates \( \widehat{\mathbf{\Theta}} \) ensures positive semi-definiteness, this approach introduces bias. To minimize this bias, truncation should instead be applied to a transformed matrix. Let \( \mathbf{G} = \int \mathbf{b}(t) \mathbf{b}(t)^{\mathrm{T}} dt \), a positive definite matrix, and define \( \widetilde{\mathbf{\Theta}} = \mathbf{G}^{1/2} \widehat{\mathbf{\Theta}} \mathbf{G}^{1/2} \). Let \( \widetilde{\mathbf{\Theta}}^* \) be the positive semi-definite truncation of \( \widetilde{\mathbf{\Theta}} \).

\begin{proposition}
\label{prop:optimal_truncation}
Let \(\widehat{C}_A(t,t') = \mathbf{b}(t)^{\mathrm{T}} \widehat{\mathbf{\Theta}} \mathbf{b}(t')\), and define the optimally truncated covariance function:
$C_A^*(t,t') = \mathbf{b}(t)^{\mathrm{T}} \mathbf{G}^{-1/2} \widetilde{\mathbf{\Theta}}^* \mathbf{G}^{-1/2} \mathbf{b}(t')$. Then \(C_A^*(t,t')\) solves the optimization problem:
$C_A^* = \operatorname*{arg\,min}_{C_A} \| \widehat{C}_A - C_A \|_{HS}, \quad \text{subject to } C_A(t,t') \text{ being positive semi-definite}$.
\end{proposition}

Proofs for Propositions 1 and 2 are provided in Supplemental Section A. Simulation studies presented in Section~\ref{sec:simul} confirm that the improved univariate test properly controls Type I error rates in sparse functional data settings. These improvements are also implemented in the multivariate test developed next.

\subsection{Multivariate Goodness-of-Fit Test of Covariance (MGFC)} \label{subsec:multi}

In this section, we extend the improved univariate test for multivariate functional data. We consider data from $N$ subjects, each with $J_i$ visits, and $K$ dependent outcomes: 
\begin{equation}
    Y_{ijk} = \mu_k(t_{ijk}) + X_{ik}(t_{ijk}) + \epsilon_{ijk}, \quad \quad 1 \leq i \leq N, 1\leq j \leq J_i, 1 \leq k \leq K,
\end{equation}
where $\mu_k(\cdot)$ is the mean function for the $k$th outcome, $\boldsymbol{X}_i(t) = (X_{i1}(t),\ldots, X_{iK}(t))'$ is $K$-variate Gaussian process with zero-mean functions and covariance functions $C^{(kk')}(t, t') = \text{cov}\{X_{ik}(t), X_{ik'}(t')\}$, and $\epsilon_{ijk}$ is Gaussian white noise independent across $i$, $j$ and $k$, with zero mean and variance $\sigma_k^2$, and independent from $\boldsymbol{X}_i$. Note that for each subject, the observation times $t_{ijk}$ may vary across outcomes.

For the null hypothesis, we assume each random function $X_{ik}(t)$ for outcome $k$ follows a linear model: $X_{ik}(t) = b_{0ik} + b_{1ik} t$, where $\boldsymbol{b}_{ik} = (b_{0ik}, b_{1ik})^{\mathrm{T}}$ are random effects. Let $\boldsymbol{b}_i = (\boldsymbol{b}_{i1}^{\mathrm{T}}, \ldots, \boldsymbol{b}_{iK}^{\mathrm{T}})^{\mathrm{T}}$ denote the concatenated random effects vector for subject $i$, which follows a multivariate normal distribution with mean zero and covariance matrix $\Sigma \in \mathbb{R}^{2K \times 2K}$. Explicit forms of the covariance functions can be derived. Specifically, if $\operatorname{var}(b_{0ik}) = \sigma_{k0}^2$, $\operatorname{var}(b_{1ik}) = \sigma_{k1}^2$, and $\operatorname{cov}(b_{0ik}, b_{1ik}) = \sigma_{k01}$, then the auto-covariance function of $X_{ik}(t)$ is $C_0^{(kk)}(t, t') = \sigma_{k0}^2 + \sigma_{k01}(t + t') + \sigma_{k1}^2 t t'$. For cross-covariances between distinct outcomes $k$ and $k'$ ($k \neq k'$), suppose $\operatorname{cov}(\boldsymbol{b}_{ik}, \boldsymbol{b}_{ik'}) = \Sigma_{kk'} \in \mathbb{R}^{2 \times 2}$. Then the cross-covariance function is given by $C_0^{(kk')}(t, t') = [1, t] \, \Sigma_{kk'} \, [1, t']^{\mathrm{T}}$. For notational simplicity, we denote the auto-covariance function as $C_0^{(k)}(t, t') := C_0^{(kk)}(t, t')$ in the remainder of the article.

The null hypothesis specifies that the auto-covariance function for each outcome equals its parametric counterpart: 
$H_0: C^{(k)}(t, t') = C_0^{(k)}(t, t')$ for all $(t, t') \in \mathcal{T}^2$ and for all $k$. 
The alternative hypothesis is that there exists some outcome $k$ and time pair $(t, t') \in \mathcal{T}^2$ such that the true covariance $C^{(k)}(t, t') \neq C_0^{(k)}(t, t')$. 
For each outcome $k$, let $\widehat{C}_0^{(k)}$ denote the estimated covariance function under the null hypothesis, and let $\widehat{C}_A^{(k)}$ denote the estimated covariance function under the alternative hypothesis, obtained using tensor-product regression splines as in the univariate test. Define the outcome-specific test statistic as $T_N^{(k)} = | \widehat{C}_A^{(k)} - \mathcal{K} \widehat{C}0^{(k)} |_{HS}$, where $\mathcal{K} \widehat{C}_0^{(k)}$ denotes the smoothed version of $\widehat{C}_0^{(k)}$ obtained via tensor-product regression splines. Based on these, we propose two aggregate test statistics:
\begin{equation}
    T_N^{\ell_\infty} = \underset{1 \leq k \leq K}{\max} T_N^{(k)},\quad
    T_N^{\ell_2} = \frac{1}{K}\sum_{k=1}^K (T_N^{(k)})^2.
    \label{eq:tnmvinf}
\end{equation}
While one could apply the univariate test separately to each $T_N^{(k)}$ with a Bonferroni correction, 
this approach ignores dependencies across outcomes and may reduce test power. 
In contrast, by pooling the individual outcome-specific statistics, the proposed approach yields a unified test that gains power through explicit incorporation of inter-outcome dependencies.

Non-rejection of the null hypothesis $H_0$ based on the proposed testing statistics suggests that the parametric covariance structure implied by the linear random intercept and slope model is adequate. Conversely, rejection of $H_0$ indicates that this model is insufficient for capturing the subject-specific trajectories in at least one outcome. In such cases, the univariate test can be subsequently applied to each outcome separately to determine which specific outcomes exhibit departures from the assumed structure, with multiple comparisons adjusted using Bonferroni's correction.

\begin{remark}[\textbf{\emph{m}-out-of-\emph{N} Bootstrap}]  
The null distribution of the proposed MGFC test statistics, as in the univariate case, is analytically intractable and requires approximation via bootstrap methods. While traditional bootstrap procedures developed by \cite{efron}, which involve resampling with the same sample size $m = N$, are widely used, they have been shown to be inconsistent for extreme value statistics such as max-type test statistics \citep{athreyabootstrapping}. In contrast, \cite{fukuchi1994bootstrapping} demonstrated that the \emph{m}-out-of-\emph{N} bootstrap, which samples $m = o(N)$ observations with replacement from the original $N$, is consistent for such distributions, provided $m/N \rightarrow 0$, $N \rightarrow \infty$, and $m \rightarrow \infty$. To ensure consistency of the bootstrap procedure, we employ traditional Efron's bootstrap ($m = N$) for $T_N^{\ell_2}$ and the \emph{m}-out-of-\emph{N} bootstrap for $T_N^{\ell_\infty}$. Specifically, we set $m = 7N^{2/3}$ for $T_N^{\ell_\infty}$, yielding bootstrap sample sizes of 100, 441, and 700 when $N = 100$, 500, and 1000, respectively. This choice satisfies the condition $m/N \to 0$, ensuring that $m = o(N)$, an asymptotic requirement in the theory of the $m$-out-of-$N$ bootstrap. This ensures that the ratio $m/N$ approaches zero as $N$ increases, a key condition for consistency. The detailed bootstrap procedure is described below.
\end{remark}

We initiate the bootstrap procedure by fitting a multivariate linear mixed model (MLMM), specified as Equation~\eqref{eq:boot}:
\begin{equation} \label{eq:boot}
y_{ijk} = \mu_k(t_{ijk}) + b_{0ik} + b_{1ik} t_{ijk} + \epsilon_{ijk}, \quad 1 \leq i \leq N, 1 \leq j \leq J_i, 1 \leq k \leq K.
\end{equation}
This MLMM includes subject-specific random intercepts and slopes for each outcome, which together capture both within-outcome temporal variability and cross-outcome dependencies. The correlation among random effects across outcomes reflects the inherent interdependence of multivariate functional data. We first remove the mean curves using the \texttt{gam} function in \texttt{R}, yielding demeaned data for each outcome. We then fit the MLMM to these demeaned outcomes to estimate the covariance matrix $\widehat{\Sigma}$, which characterizes the covariance among the full vector of random effects $(b_{0i1}, b_{1i1}, b_{0i2}, b_{1i2}, \ldots, b_{0iK}, b_{1iK})$, resulting in a $2K \times 2K$ matrix.

Given $\widehat{\Sigma}$, we generate bootstrap samples under the null hypothesis using a parametric approach. Specifically, for each bootstrap replicate $b$, we draw random effects from $\mathcal{N}(0, \widehat{\Sigma})$ to construct the bootstrap random functions $X_{ik}^{(b)}(t_{ijk}) = \widehat{b}_{0ik}^{(b)} + \widehat{b}_{1ik}^{(b)} t_{ijk}$ for $i = 1, \ldots, m$, where $m = N$ for the $T_N^{\ell_2}$ statistic and $m = 7N^{2/3}$ for $T_N^{\ell_\infty}$. We then generate bootstrap observations $Y_{ijk}^{(b)}$ by adding Gaussian noise to $X_{ik}^{(b)}(t_{ijk})$, where the noise variance $\sigma_k^2$ is estimated using FACEs under the alternative model, as described in Section~\ref{subsec:face}.

For each bootstrap replicate, we compute test statistics $T_{mb}^{(k)}$ for each outcome $k = 1, \ldots, K$, forming vectors $\mathbf{T}_m^{(1)*}, \ldots, \mathbf{T}_m^{(K)*}$, where $\mathbf{T}_m^{(k)*} = (T_{m1}^{(k)}, \ldots, T_{mB}^{(k)})'$ represents the bootstrap distribution of $T_N^{(k)}$. These vectors are assembled into a $B \times K$ matrix $M_m$, with column $k$ corresponding to $\mathbf{T}_m^{(k)*}$. For each row of $M_m$, we compute the maximum value across columns to obtain $T_{m1}^{\ell_\infty}, \ldots, T_{mB}^{\ell_\infty}$, which approximate the null distribution of $T_N^{\ell_\infty}$. The null distribution of $T_N^{\ell_2}$ is similarly approximated using Efron's bootstrap with $m = N$. Corresponding p-values are derived from these bootstrap distributions. Algorithm~\ref{alg:bootstrap} summarizes the full procedure for $T_N^{\ell_\infty}$.

\setlist[itemize]{noitemsep, topsep=0pt}
\setlist[enumerate]{noitemsep, topsep=0pt}
\RestyleAlgo{ruled}
\begin{algorithm}
\caption{Parametric bootstrap algorithm for estimating the null distributions of $T_N^{\ell_\infty}$ and $T_N^{\ell_2}$}
\label{alg:bootstrap}
\begin{enumerate}
\item Fit the multivariate linear mixed model in Equation~\eqref{eq:boot} to the demeaned data and estimate the covariance matrix $\widehat{\Sigma}$ for the random effects vector $(b_{0i1}, b_{1i1}, \ldots, b_{0iK}, b_{1iK})$.

\item Fit FACEs to each of the demeaned data for outcome $k$ and estimate the error variance $\widehat{\sigma}_k^2$ (Section~\ref{subsec:face}).

\item \For{$b = 1, \ldots, B$ bootstrap replicates}{
\begin{enumerate}
\item Sample random effects $(\widehat{b}_{0i1}^{(b)}, \widehat{b}_{1i1}^{(b)}, \ldots, \widehat{b}_{0iK}^{(b)}, \widehat{b}_{1iK}^{(b)})$ from $\mathcal{N}(0, \widehat{\Sigma})$.

\item \For{$k = 1, \ldots, K$ outcomes}{
\begin{enumerate}
\item Generate $X_{ik}^{(b)}(t_{ijk}) = \widehat{b}_{0ik}^{(b)} + \widehat{b}_{1ik}^{(b)} t_{ijk}$ for $i = 1, \ldots, m$, where $m = o(N)$ for $T_N^{\ell_\infty}$ and $m = N$ for $T_N^{\ell_2}$.

\item Generate errors $\epsilon_{ijk}^{(b)} \sim \mathcal{N}(0, \widehat{\sigma}_k^2)$, with $\widehat{\sigma}_k^2$ estimated via FACEs.

\item Form bootstrap observations $Y_{ijk}^{(b)} = \widehat{\mu}(t_{ijk}) + X_{ik}^{(b)}(t_{ijk}) + \epsilon_{ijk}^{(b)}$.

\item Compute $T_{mb}^{(k)} = \lVert \widehat{C}_A^{(b)} - \mathcal{K} \widehat{C}_0^{(b)} \rVert_{HS}$ using the sample size $m$.
\end{enumerate}
}

\item Calculate aggregate bootstrap statistics:
\[
T_{mb}^{\ell_\infty} = \max_{1 \leq k \leq K} T_{mb}^{(k)}, \quad T_{mb}^{\ell_2} = \frac{1}{K} \sum_{k=1}^K \left(T_{mb}^{(k)}\right)^2.
\]
\end{enumerate}
}

\item Compute p-values using the bootstrap samples:
\begin{itemize}
    \item[-] For $T_N^{\ell_\infty}$: $p = B^{-1} \sum_{b=1}^B \mathbb{I}(T_{mb}^{\ell_\infty} > T_N^{\ell_\infty})$.
    \item[-] For $T_N^{\ell_2}$: $p = B^{-1} \sum_{b=1}^B \mathbb{I}(T_{mb}^{\ell_2} > T_N^{\ell_2})$.
\end{itemize}
\end{enumerate}
\end{algorithm}

\section{Simulation} \label{sec:simul}

We conduct simulation studies to evaluate the performance of the proposed tests in both univariate and multivariate settings. Specifically, Section~\ref{sec:uniImproved} assesses Type I error control and power for our improved univariate goodness-of-fit test, comparing it with the original univariate test \citep{chen2019smoothing}. Section~\ref{sec:multiImproved} evaluates the proposed multivariate goodness-of-fit test (MGFC) in terms of Type I error control and power under two designs: a balanced sparse design in Section~\ref{sec:sim1}, and an unbalanced highly sparse design informed by real data in Section~\ref{sec:sim2}.

\subsection{Simulation Study: Improved Univariate Goodness-of-Fit Test}
\label{sec:uniImproved}
In this subsection, we evaluate the performance of the improved univariate test and compare it with the original univariate test of \citet{chen2019smoothing}. Data are generated under the null hypothesis from a random effects model with linear random intercept and slope, implying a quadratic polynomial covariance structure with $\sigma_0^2 = 1$ and $\sigma_{01} = -0.5$. The mean function is set to $\mu(t) = 0$. We consider sample sizes $N \in {100, 500}$ and three values for the error variance $\sigma^2 \in {0.25, 1, 4}$ to assess performance across varying noise levels. For each subject, candidate time points $t_{ij}$ are selected from 80 equally spaced points on $[-1,1]$, with subject-specific observation times randomly sampled to simulate sparsity. The number of observations per subject is sampled from ${2,3,4,5,6}$ for an average of $\bar{J} = 4$ and from ${5,6,7,8,9}$ for $\bar{J} = 7$, representing highly and moderately sparse settings, respectively. For instance, $\bar{J} = 4$ corresponds to fewer observations per subject, mimicking real-world sparsity.

We compute empirical Type I error rates as the proportion of rejections under the null hypothesis at nominal levels $\alpha \in {0.05, 0.10}$, based on 5000 replications. A valid method should yield empirical Type I errors close to the nominal $\alpha$. Table~\ref{tab:type1_error} summarizes empirical Type I error rates, highlighting performance differences between the two tests. The original univariate test shows substantial inflation in Type I error, especially under high sparsity and large error variance. In contrast, the improved test consistently achieves error rates much closer to the nominal levels across all settings, confirming its robustness in sparse functional data settings. Additional simulation results comparing power curves under two alternative scenarios are provided in Supplement Section B.

\begin{table}[htbp]
\centering
\caption{Empirical Type I error rates for the original univariate test and the improved test, based on 5000 simulation replications across varying sample sizes ($N$), error variances ($\sigma^2$), and average visit numbers ($\bar{J}$). The nominal significance level $\alpha$ denotes the target Type I error rate, set at 0.05 and 0.10.}
\begin{tabular*}{0.9\textwidth}{@{\extracolsep{\fill}}lcccccc}
\toprule
 & &  & \multicolumn{2}{c}{Original Univariate Test} & \multicolumn{2}{c}{Improved Univariate Test} \\ 
\cmidrule(lr){4-5} \cmidrule(lr){6-7}
\textbf{$\sigma^2$} & $N$ & $\bar{J}$ & $\alpha = 0.05$ & $\alpha = 0.10$ & $\alpha = 0.05$ & $\alpha = 0.10$ \\ 
\midrule
0.25 & 100 & 4 & 0.062 & 0.120 & 0.054 & 0.103 \\
     &     & 7 & 0.053 & 0.112 & 0.047 & 0.089 \\
     & 500 & 4 & 0.063 & 0.115 & 0.054 & 0.097 \\
     &     & 7 & 0.051 & 0.107 & 0.043 & 0.093 \\
\midrule
1    & 100 & 4 & 0.100 & 0.191 & 0.052 & 0.103 \\
     &     & 7 & 0.071 & 0.146 & 0.044 & 0.089 \\
     & 500 & 4 & 0.081 & 0.154 & 0.051 & 0.099 \\
     &     & 7 & 0.073 & 0.132 & 0.041 & 0.094 \\
\midrule
4    & 100 & 4 & 0.202 & 0.318 & 0.047 & 0.101 \\
     &     & 7 & 0.136 & 0.237 & 0.040 & 0.091 \\
     & 500 & 4 & 0.151 & 0.244 & 0.053 & 0.096 \\
     &     & 7 & 0.093 & 0.169 & 0.047 & 0.095 \\
\bottomrule
\end{tabular*}
\label{tab:type1_error}
\end{table}

\subsection{Simulation Study: Multivariate Goodness-of-Fit Test}
\label{sec:multiImproved}

We evaluate the performance of the proposed MGFC test in terms of Type I error control and power, comparing it with the adjusted univariate test $T_N^B$. We assess both the $\ell_\infty$ and $\ell_2$ versions of the MGFC test, denoted $T_N^{\ell_\infty}$ and $T_N^{\ell_2}$, respectively, and compare them with the improved univariate test applied individually to each outcome. For the latter, we apply Bonferroni correction to account for multiple testing and denote the resulting adjusted procedure as $T_N^B$, with superscript $B$ indicating Bonferroni adjustment.

Two simulation scenarios are considered. In Section~\ref{sec:sim1}, we examine a balanced design with three dependent outcomes ($K=3$), where all subjects share common, regularly spaced observation times. Section~\ref{sec:sim2} explores an unbalanced, highly sparse design involving two outcomes ($K=2$), with subject-specific visit schedules mimicking those observed in the ADNI dataset. This design reflects real-world heterogeneity in visit patterns, offering a practical evaluation of the MGFC test performance under realistic sparsity.

\subsubsection{Scenario 1: Balanced Design with Three Outcomes} \label{sec:sim1}

We evaluate the MGFC test under a balanced design with three dependent outcomes. Data are generated according to the model:
\begin{align}
\label{eq:simeq1}
    \nonumber Y_{ijk} &= \mu(t_{ij}) + X_{ik}(t_{ij}) + \epsilon_{ijk}, \\
    X_{ik}(t_{ij}) &= b_{0ik} + b_{1ik} t_{ij} + \Delta_k z_i(t_{ij}), \quad 1 \leq i \leq N,\; 1 \leq j \leq J_i,\; 1 \leq k \leq K,
\end{align}
where the effect size parameter $\Delta_k$ controls the magnitude of deviation from the null model through the deviation function $z_i(t_{ij})$. All outcomes share common time points, i.e., $t_{ijk} = t_{ij}$ for all $k$. The mean function is set to $\mu(t) = 0$, and the residuals $\epsilon_{ijk} \sim \mathcal{N}(0,1)$ are independent of $X_{ik}$. The vector of random intercepts and slopes, $(b_{0i1}, b_{1i1}, b_{0i2}, b_{1i2}, b_{0i3}, b_{1i3})$, follows a 6-dimensional multivariate normal distribution with mean zero and a positive definite covariance matrix, which induces correlation among the three outcomes. Details of the covariance structure are provided in Supplement Section C.

All subjects share the same number of observations, denoted by $J$, with observation times fixed across subjects and outcomes, thereby ensuring a balanced design across individuals and outcomes. Simulations are conducted across a range of sample sizes ($N$) and visit numbers ($J$), with each configuration replicated 1000 times. To assess Type I error control, we set $\Delta_k = 0$ for all $k$, thereby generating data under the null hypothesis. Table~\ref{table:type1dependent} presents the empirical Type I error rates for $T_N^{\ell_\infty}$ and $T_N^{\ell_2}$ across various combinations of $N$ and $J$, at nominal significance levels $\alpha = 0.05$ and 0.10. The observed rates closely align with the nominal levels, indicating robust Type I error control, even when $N = 100$, under both test statistics.

\begin{table}[htbp]
\centering
\begin{tabular}{ccccc}
\hline 
 &  \multicolumn{2}{c}{$T_N^{\ell_\infty}$}  &  \multicolumn{2}{c}{$T_N^{\ell_2}$} \\
\hline
 &  $\alpha = 0.05$ & $\alpha = 0.10$ & $\alpha = 0.05$ & $\alpha = 0.10$ \\
\hline
N=100, J=5  & 0.030 & 0.106  & 0.049 & 0.107 \\
N=100, J=10 & 0.056 & 0.096  & 0.058 & 0.103 \\
N=100, J=15 & 0.039 & 0.104  & 0.045 & 0.112 \\
N=100, J=20 & 0.043 & 0.097  & 0.047 & 0.098 \\
N=500, J=5 & 0.051 & 0.098  & 0.055 & 0.097 \\
N=500, J=10 & 0.051 & 0.102  & 0.056 & 0.113 \\
N=500, J=15 & 0.047 & 0.096  & 0.052 & 0.092 \\
N=500, J=20 & 0.048 & 0.099  & 0.045 & 0.102 \\
\hline
\end{tabular}
\caption{Empirical Type I error rates for the MGFC test statistics $T_N^{\ell_\infty}$ and $T_N^{\ell_2}$, based on 1000 simulation replications under the null hypothesis. Results are shown for varying sample sizes ($N$) and visit numbers per subject ($J$), at nominal significance levels $\alpha = 0.05$ and 0.10.} 
\label{table:type1dependent}
\end{table}

For the power analysis, we consider two forms of deviation from the null hypothesis: a quadratic deviation and a trigonometric deviation, defined as follows:
\begin{enumerate}[label = (\Roman*)]
    \item Quadratic deviation: $z_i(t) = b_{2i} t^2$, where $b_{2i} \sim \mathcal{N}(0,1)$, \label{quadDev}
    \item Trigonometric deviation: $z_i(t) =  \xi_{i1}\sin(2\pi t) + \xi_{i2} \sin(4 \pi t)$, where $\xi_{i1}, \xi_{i2} \sim \mathcal{N}(0,1)$. \label{trigDev}
\end{enumerate}

As specified in Equation~\eqref{eq:simeq1}, the effect size parameter $\Delta_k$ modulates the influence of the deviation function $z_i(t)$ for each outcome $k$. To evaluate sensitivity across different effect patterns, we consider two configurations. In Scenario 1(a), we set $\Delta_k = \Delta$ for all $k$, corresponding to equal deviations across outcomes. In Scenario 1(b), we set $\Delta_1 = \Delta$ and $\Delta_2 = \Delta_3 = 0$, so that only the first outcome exhibits deviation from the null model. In both configurations, the effect size $\Delta$ ranges from 0.2 to 1.5 to generate power curves. We evaluate each scenario under two levels of visit numbers per subject: $J = 5$ and $J = 10$.

\begin{figure}[htbp]
    \centering
    \includegraphics[width = 0.9\textwidth]{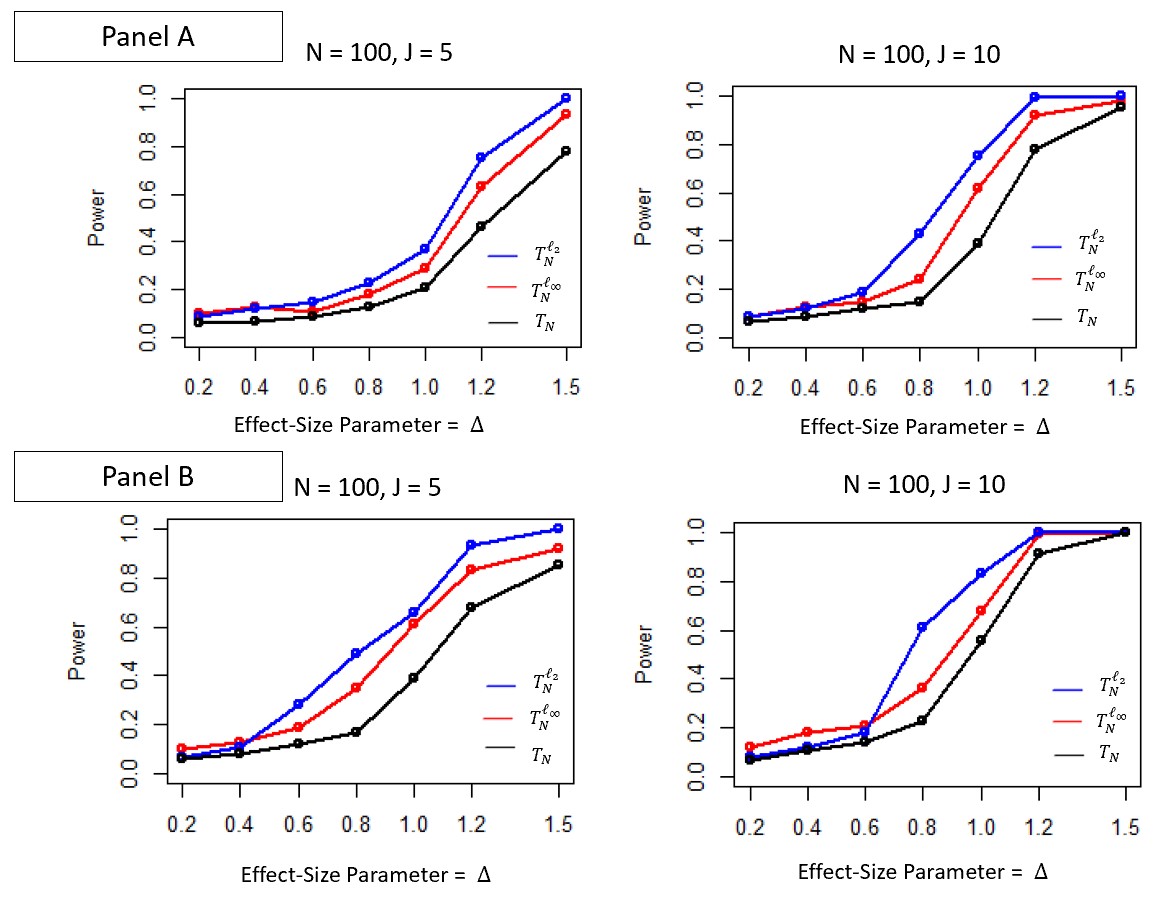}
    \caption{Scenario 1(a): Empirical power curves for the MGFC test in $\ell_2$ (blue) and $\ell_\infty$ (red) forms, denoted $T_N^{\ell_2}$ and $T_N^{\ell_\infty}$, respectively, compared with the adjusted univariate test $T_N^B$ (black). The effect size $\Delta$ is applied equally to all outcomes. Panel A shows results for the quadratic deviation function~\ref{quadDev}, and Panel B for the trigonometric deviation function~\ref{trigDev}, across visit numbers $J = 5$ and $10$.} \label{fig:simul_Scenario1}
\end{figure}

\begin{figure}[htbp]
    \centering
    \includegraphics[width = \textwidth]{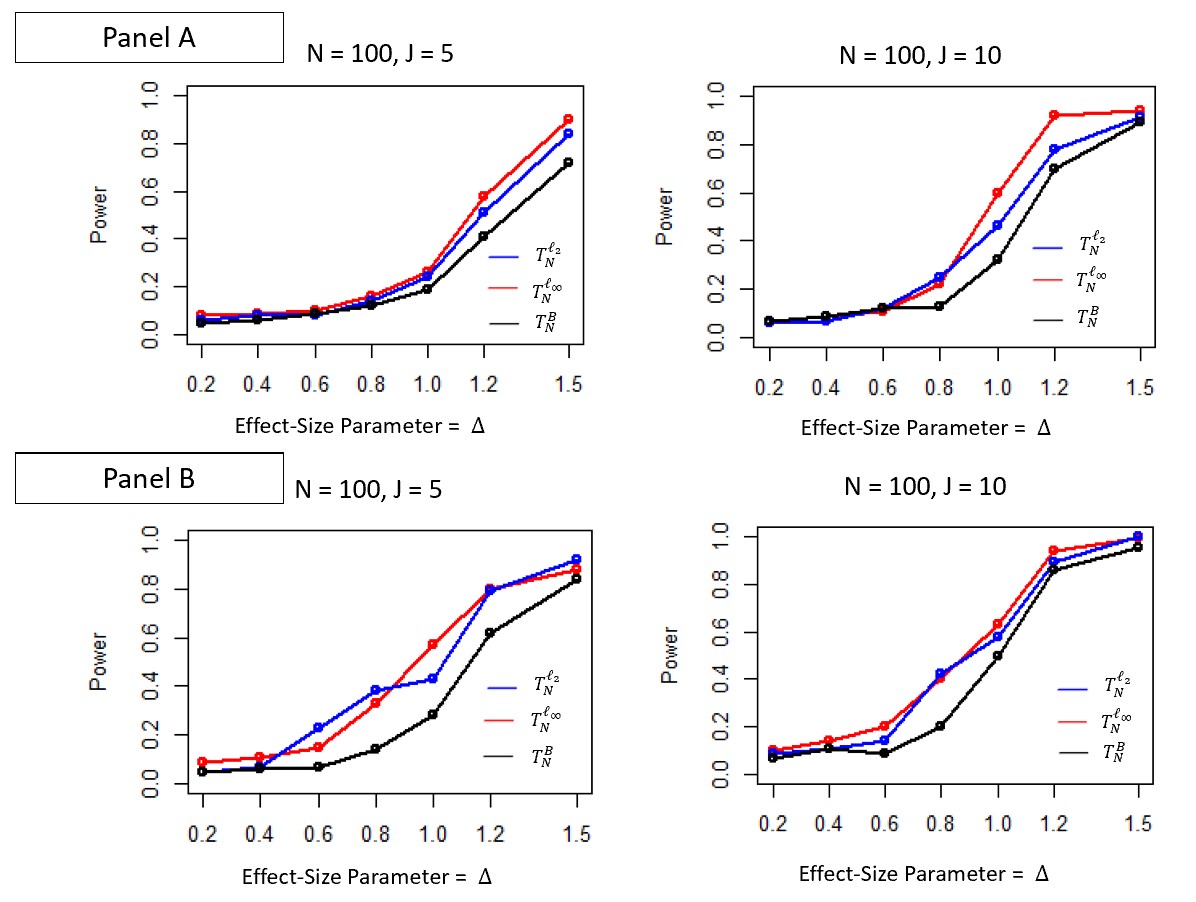}
    \caption{Scenario 1(b): Empirical power curves for $T_N^{\ell_2}$ (blue), $T_N^{\ell_\infty}$ (red), and the adjusted univariate test $T_N^B$ (black), where only the first outcome has nonzero effect size $\Delta$. Panel A corresponds to the quadratic deviation function~\ref{quadDev}, and Panel B to the trigonometric deviation function~\ref{trigDev}, across visit numbers $J = 5$ and $10$.}
 \label{fig:simul_Scenario1setting2}
\end{figure}

Figure~\ref{fig:simul_Scenario1} and Figure~\ref{fig:simul_Scenario1setting2} display empirical power curves for the two forms of the MGFC test and the adjusted univariate test $T_N^B$ across varying effect sizes $\Delta$ and visit numbers $J$ under Scenarios 1(a) and 1(b), respectively. As expected, power increases with larger $\Delta$. Across all settings, both forms of the MGFC test consistently outperform the adjusted univariate test, demonstrating greater sensitivity to deviations from the specified parametric covariance structure. In Scenario 1(a), where all outcomes deviate equally, the $\ell_2$ form ($T_N^{\ell_2}$) achieves higher power than the $\ell_\infty$ form ($T_N^{\ell_\infty}$). In contrast, when only one outcome deviates (Scenario 1(b)), $T_N^{\ell_\infty}$ yields greater power. These findings suggest that the choice between the $\ell_2$ and $\ell_\infty$ forms may depend on the sparsity and distribution of effects across outcomes, though both variants of the MGFC test consistently outperform $T_N^B$.

\subsubsection{Scenario 2: Unbalanced Design with Two Outcomes} \label{sec:sim2}

Section~\ref{sec:sim1} focused on a balanced design with regularly spaced time points and a constant number of observations per subject. In contrast, real-world longitudinal studies often involve substantial variability in observation frequency, with many subjects having limited follow-up. For instance, while \citet{chen2019smoothing} evaluated the univariate test using a minimum of $J = 10$ observations per subject, this assumption assumes a higher visit frequency than is typical in real-world studies, where most subjects have substantially fewer visits. In the ADNI study, for example, over 70\% of subjects have fewer than 5 observations. As such, power evaluations based on $J = 10$ may overstate performance in practice. We therefore adopt the improved univariate test from Section~\ref{sec:uniImproved}, which accommodates sparse data with fewer visits per subject.

To simulate a realistic unbalanced design, we use the ADNI dataset to emulate the empirical distribution of visit frequencies, as detailed in Section D. On average, subjects attended approximately 6 visits, with the majority having fewer than 5. Based on this distribution, we simulate data for $N = 500$ subjects and replicate the simulation 1000 times to ensure reliable inference.

For Type I error evaluation, we generate data under the null hypothesis by setting $\Delta_k = 0$ for all $k$. The empirical Type I error rates for $T_N^{\ell_\infty}$ were 0.053 at $\alpha = 0.05$ and 0.111 at $\alpha = 0.10$, while for $T_N^{\ell_2}$, they were 0.051 and 0.099, respectively. These results confirm satisfactory Type I error control for both forms of the MGFC test, even under substantial visit imbalance.

\begin{figure}[htbp]
    \centering
    \includegraphics[width = \textwidth]{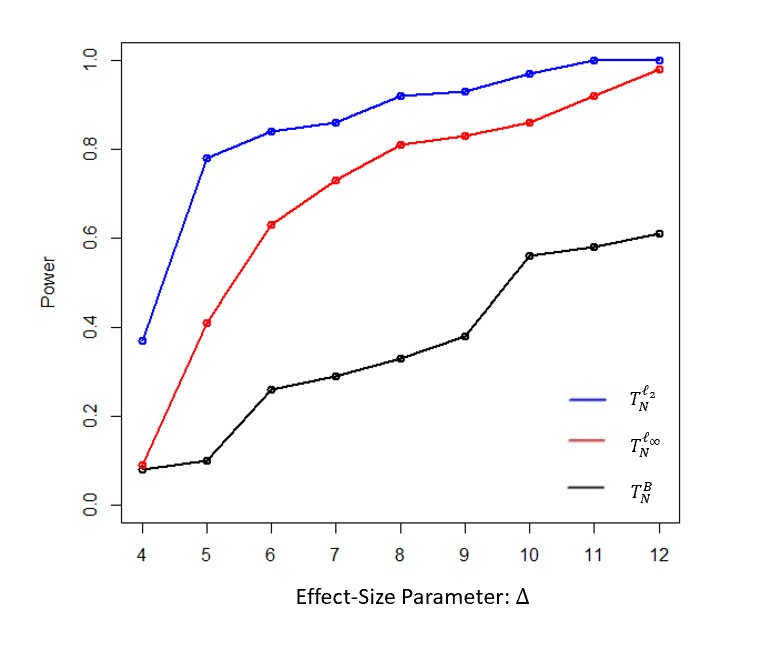}
    \caption{Scenario 2: Empirical power curves for the MGFC test in $\ell_2$ (blue) and $\ell_\infty$ (red) forms, denoted $T_N^{\ell_2}$ and $T_N^{\ell_\infty}$, respectively, compared with the adjusted univariate test $T_N^B$ (black). The effect size $\Delta$ is applied equally to both outcomes under the quadratic deviation function~(I) defined in Scenario 1.} 
    \label{fig:simul_Scenario2}
\end{figure}

We next assess power under the quadratic deviation function~(I), with effect sizes $\Delta_k = \Delta$ for all $k$. We vary $\Delta$ from 4 to 12 to generate power curves across a range of effect magnitudes. Figure~\ref{fig:simul_Scenario2} displays the empirical power curves for $T_N^{\ell_\infty}$, $T_N^{\ell_2}$, and the adjusted univariate test $T_N^B$. As expected, power increases with $\Delta$, and both forms of the MGFC test consistently outperform the adjusted univariate test across the entire range, demonstrating improved detection of covariance misspecification under realistic sparsity patterns.

\section{Applications to Real-World Longitudinal Data}
\label{sec:realData}

\subsection{ADNI Data: Covariance Structure Assessment in Longitudinal Outcomes} \label{sec:adni}
We analyze data from the Alzheimer's Disease Neuroimaging Initiative (ADNI) \citep{petersen2010alzheimer}, a landmark longitudinal study designed to discover and validate biomarkers for the early detection and tracking of Alzheimer's disease (AD). ADNI collects extensive longitudinal data through cognitive tests, biomarker assays, and neuroimaging techniques to better understand AD progression. For our analysis, we focused on five key outcomes ($K = 5$) previously used by \cite{lin2021functional} to model cognitive decline: the Alzheimer's Disease Assessment Scale-Cognitive 13 items (ADAS-Cog 13), Functional Assessment Questionnaire (FAQ), Rey Auditory Verbal Learning Test (RAVLT) immediate and RAVLT learning scores, and the Mini-Mental State Examination (MMSE). These outcomes assess multiple domains of cognitive and functional ability impacted by AD. To facilitate a unified analysis, we applied transformations so that higher values consistently indicated better outcomes across all measures. The LOWESS curves for visualizing longitudinal trends of these outcomes in the MCI group are shown in Supplemental Section D.

Subjects with at least two visits and complete data for all five variables were included. We analyzed three subgroups based on baseline clinical status: mild cognitive impairment (MCI, $n=849$), dementia ($n=556$), and cognitively normal (CN, $n=554$). To assess whether a linear covariance structure adequately described the joint distribution of outcomes in each subgroup, we applied the multivariate goodness-of-fit test (MGFC) in both $\ell_\infty$ and $\ell_2$ forms, using 1000 bootstrap replications to estimate p-values. The adjusted univariate test, $T_N^B$, was also computed for each outcome with Bonferroni correction ($\alpha = 0.01$) to control for multiple testing across $K = 5$ outcomes. The resulting test statistics and p-values are summarized in Table~\ref{tab:adni}.

In the MCI subgroup, the MGFC test yielded $T_N^{\ell_\infty} = 7.69$ ($p < 0.001$) and $T_N^{\ell_2} = 21.18$ ($p < 0.001$), indicating evidence of nonlinearity in the covariance structure. Univariate testing identified FAQ as the only outcome with evidence of nonlinearity ($p < 0.001$, $T_N = 6.899$), below the Bonferroni threshold of 0.01. In the dementia subgroup, $T_N^{\ell_\infty} = 6.29$ ($p = 0.053$) and $T_N^{\ell_2} = 20.70$ ($p = 0.037$), suggesting marginal significance. Univariate tests did not identify any outcomes with p-values below 0.01, indicating insufficient evidence of nonlinearity in individual outcomes, despite marginal evidence at the multivariate level. In the CN subgroup, $T_N^{\ell_\infty} = 3.74$ ($p = 0.412$) and $T_N^{\ell_2} = 6.14$ ($p = 0.641$), indicating no deviation from linearity. These findings highlight the utility of the MGFC test in detecting nonlinearity in multivariate longitudinal data, particularly in the MCI group, where FAQ exhibited significant nonlinear covariance structure.

\begin{table}[htbp]
\centering
\begin{tabular}{l c c c c c c}
\hline
Subgroup & $T_N^{\ell_\infty}$  & p-value & $T_N^{\ell_2}$ & p-value & $\max_{k} T_N^k$ & min p-value  \\
\hline
MCI ($n = 849$)     & $7.69$  & $<0.001$  & $21.18$ & $<0.001$ & $6.899$ & $<0.001$ \\
Dementia ($n = 556$) & $6.29$ & $0.053$     & $20.70$ & $0.037$    & $6.102$ & $0.035$    \\
CN ($n = 554$)      & $3.74$ & $0.412$     & $6.14$  & $0.641$    & $3.829$ & $0.203$    \\
\hline
\end{tabular}
\caption{Test statistics and p-values for ADNI subgroups using the MGFC test in $\ell_\infty$ and $\ell_2$ forms, and the adjusted univariate test. The last two columns report the largest univariate test statistic $\max_{k} T_N^k$ and the smallest p-value across the five outcomes, compared to a Bonferroni-adjusted significance level of $0.01$ (i.e., $0.05 / 5$).}
\label{tab:adni}
\end{table}

\subsection{PPMI Data: Covariance Structure Assessment in Parkinson's Disease Outcomes} \label{sec:parkinsons}

We analyze data from the Parkinson's Progression Marker Initiative (PPMI) \citep{marek2011parkinson}, a longitudinal, multi-center observational study designed to characterize clinical progression and identify biomarkers across all stages of Parkinson's disease (PD), from prodromal to moderate disease. The study collects extensive longitudinal data, including clinical, imaging, biological, and digital measures. Our analysis focuses on the Movement Disorder Society Unified Parkinson's Disease Rating Scale (MDS-UPDRS) \citep{Goetz2008MDS}, which evaluates Parkinsonian symptoms and their impact on daily living. The MDS-UPDRS comprises 65 items across four parts, each rated on a 5-point scale (0–4, higher scores indicating greater severity). We consider three primary outcomes derived from this scale ($K = 3$): sum scores for Part I (Non-Motor Aspects of Experiences of Daily Living; range 0–52), Part II (Motor Aspects of Experiences of Daily Living; range 0–52), and Part III (Motor Examination; range 0–132). The LOWESS curves for visualizing longitudinal trends of these outcomes are provided in Supplemental Section E.

We include subjects diagnosed with PD at baseline and with at least two visits, yielding a sample size of $N = 785$. We apply the MGFC test in both $\ell_\infty$ and $\ell_2$ forms, using 1000 bootstrap replications to compute p-values, and assess nonlinearity in covariance structures. Both $T_N^{\ell_\infty}$ and $T_N^{\ell_2}$ yield p-values below 0.001, indicating strong evidence of deviation from a linear covariance structure. We further conduct univariate tests for each outcome, all yielding p-values less than 0.001, which remain significant under a Bonferroni-adjusted threshold of 0.017 (i.e., $0.05/3$).

These findings reveal significant nonlinearity in the covariance structures of all three MDS-UPDRS outcomes, highlighting the complexity of PD progression and the importance of accurately modeling these deviations in multivariate longitudinal analyses.

\section{Discussion}
\label{sec:concl}
We proposed a multivariate goodness-of-fit test for assessing covariance structures in longitudinal data, called the Multivariate Goodness-of-Fit Covariance (MGFC) test. Building on the univariate test developed by \citet{chen2019smoothing}, we introduced an improved version that ensures valid Type I error control under sparse designs. This enhanced univariate test was further extended to the multivariate context using test statistics based on the $\ell_\infty$ and $\ell_2$ norms, allowing for a unified assessment of whether linear mixed-effects models adequately capture the underlying covariance structure across multiple outcomes. Extensive simulations demonstrated that the MGFC test maintains proper Type I error control and achieves higher power than univariate tests adjusted for multiple comparisons, especially under sparse or unbalanced designs.

We applied the MGFC tests to two longitudinal datasets: ADNI and PPMI. In the ADNI dataset, we found evidence of nonlinearity in the covariance structure of the Functional Assessment Questionnaire (FAQ) within the mild cognitive impairment subgroup, while linear covariance structures were adequate for the dementia and cognitively normal subgroups. In the PPMI dataset, MGFC detected significant nonlinearity across all three outcomes derived from the MDS-UPDRS scale, reflecting the complexity of Parkinson’s disease progression. These applications illustrate the value of the MGFC test in detecting nonlinear covariance patterns not identified by univariate analyses, thereby enhancing model assessment in real-world longitudinal studies.

Future work may extend this framework to accommodate non-Gaussian outcomes or high-dimensional data and explore its integration with machine learning methods for more flexible modeling. By offering a robust and scalable method for assessing covariance structure in multivariate longitudinal data, the MGFC test supports more accurate modeling of disease progression and contributes to the development of diagnostic and therapeutic strategies grounded in a deeper understanding of temporal dependencies among clinical outcomes.


\section*{Acknowledgements}

Alzheimer's Data used in the preparation of this article (Section \ref{sec:adni}) were obtained from the Alzheimer's Disease Neuroimaging Initiative (ADNI) database (\href{adni.loni.usc.edu}{adni.loni.usc.edu}). As such, the investigators within the ADNI contributed to the design and implementation of ADNI and/or provided data but did not participate in the analysis or writing of this report. A complete listing of ADNI investigators can be found at \href{http://adni.loni.usc.edu/wp-content/uploads/how_to_apply/ADNI_Acknowledgement_List.pdf}{http://adni.loni.usc.edu
}.

Parkinson's Data used in the preparation of this article (Section \ref{sec:parkinsons}) were obtained from the Parkinson's Progression
Markers Initiative (PPMI) database (\href{www.ppmi-info.org/access-data-specimens/download-data}{www.ppmi-info.org/access-data-specimens/download-data}).
For up-to-date information on the study, visit \href{www.ppmi-info.org}{www.ppmi-info.org}.
\vspace*{-8pt}


\bibliography{biomsample}
\bibliographystyle{biom}

\end{document}